# Uncorrelated two-state single molecule trajectories from reducible kinetic schemes*

Ophir Flomenbom, and Joseph Klafter

*School of Chemistry, Raymond & Beverly Sackler Faculty of Exact Sciences, Tel Aviv University, Ramat Aviv, Tel Aviv 69978, Israel*

**Abstract**

Trajectories of *on-off* events are the output of many single molecule experiments. Usually, one describes the underlying mechanism that generates the trajectory using a kinetic scheme, and by analyzing the trajectory aims at deducing this scheme. In a previous work [O. Flomenbom, J. Klafter, and A. Szabo, submitted (2004)], we showed that when successive events along a trajectory are uncorrelated, all the information in the trajectory is contained in two basic functions, which are the waiting time probability functions (PDFs) of the *on* state and of the *off* state. The kinetic schemes that lead to such uncorrelated trajectories were termed *reducible*. Here we discuss the reasons that lead to reducible schemes. In particular, the topology of reducible schemes is characterized and proven.

## Introduction

A large number of complex systems have been recently studied both experimentally[1-27] and theoretically[28-43] on the single molecule level. Examples include the flux of ions through individual channels[1, 21-24], the translocation of ssDNA and RNA through single nanopores[25-26], diffusion of single molecules[4-8], conformational fluctuation of biopolymers[9-15], single enzyme activity[16-20], and nano-crystals blinking[27]. Many of these processes are characterized by multi-substate kinetic schemes with time independent transition probabilities, where the corresponding dynamics are described either by the master equation[44], or by the generalized master equation[43]. Due to the complexity of the system, it is not possible in most cases to observe all its substates. Usually the observable at time *t* fluctuates between two distinct values, implying that each of the substates belongs to one of two possible states: *on* state and *off* state. This produces a time-series of *on-off* events, which is called a trajectory (Fig. 1). From the analysis of the two-state trajectory one wishes to determine the underlying multi-substate kinetic scheme which generates it. Generally, in a multi-substate scheme, the number of substates in each of the states can vary (Figs. 2A-2B), and the connectivity between substates within a state and between states can be complex, namely, it can exceed the one dimensional nearest neighbors connectivity within a state (Fig. 2B) and can contain a complex network of connections between substates of different states (Figs. 2C-2E). Additionally, the schemes may show a net flow in steady state along some connections (i. e. a non-equilibrium steady state), when an external source of energy is present[45].

The basic characterization of the time-series is given by the waiting time PDFs of the *on* state, $\phi_{on}(t)$, and of the *off* state, $\phi_{off}(t)$. $\phi_{on}(t)$ and $\phi_{off}(t)$ cannot be obtained, in principle, from bulk measurements, but are easily obtained from the single molecule trajectory by building histograms from the random time durations that the observable occupies the *on*-state and *off*-state, respectively. Given $\phi_{on}(t)$ and $\phi_{off}(t)$ from the experiment, one aims at adjusting the details of a multi-substate scheme so that the calculated $\phi_{on}(t)$ and $\phi_{off}(t)$ are compatible with the experimental ones.

$\phi_z(t)$ ($z = on, off$) is calculated from the expression,

$$\phi_z(t) = \sum_{n,m} W_n^z f_{mn}^z(t) = \sum_n W_n^z F_n^z(t). \qquad (1)$$

Here, $W_n^z$ is the probability that a given event in state $z$ starts at substate $n$ of this state, $f_{mn}^z(t)$ is the conditional waiting time PDF to start an event at substate $n$ of state $z$ and to terminate it at substate $m$ of the other state, and $F_n^z(t) = \sum_m f_{mn}^z(t)$ is the conditional waiting time PDF to start an event at substate $n$ of state $z$ and terminate it at any substate of the other state. Note that $W_n^z$ is the normalized steady state flux from the other state to substate $n$ of state $z$, and is found by solving the (reversible) coupled (*on-off*) dynamic equation (either the master equation or the generalized master equation) in steady state, where $f_{mn}^z(t)$ is found from the appropriate Green function of the (irreversible) decoupled dynamic equation for state $z$.

Given a kinetic scheme, $\phi_{on}(t)$ and $\phi_{off}(t)$ are found from Eq. (1), and, thus, can be made the same as the experimentally obtained waiting time PDFs. However, when these functions are multi-exponentials, there are several kinetic schemes that lead to the same waiting time PDFs. The question is, therefore: *can one discriminate between kinetic schemes that lead to the same waiting time PDFs?* More specifically, one can inquire if other functions calculated from the trajectory can supply additional information useful in discriminating among schemes with the same waiting time PDFs. Such functions include: (*a*) the two successive waiting times PDFs[16, 21, 32-34, 36, 42], $\phi_{x,y}(t_1,t_2)$  $x,y = on, off$, (*b*) the *on-off* propagator[11, 16-20, 31-31, 34, 36-40], $G(t_2\ x\,|\,t_1\ y) = G(t\ x\,|\,0\ y)$, which is the bulk relaxation function, where the equality is valid for stationary processes as we consider here, (*c*) higher order propagators[18, 31, 34, 36-37, 43], e. g. $G(t_2\ x; t_1\ y\,|\,0\ z)$, $x,y,z = on, off$, and (*d*) higher order successive waiting times PDFs[36], e. g. $\phi_{x,y,z}(t_1,t_2,t_3)$. Note that the functions (*a*), (*c*) and (*d*) can be obtained only from single molecule trajectories. When events along the trajectory are uncorrelated, all the information in the trajectory is contained in the waiting time PDFs. We call those schemes that lead to uncorrelated trajectories *reducible*[36]. It follows that reducible schemes with the same waiting time PDFs cannot be distinguished from each other by analyzing a trajectory. Our main focus in this paper is on the reasons that lead to reducible schemes. In particular, we characterize and prove the topology of reducible kinetic schemes.

**Reducible Schemes**

Consider a trajectory with no correlations between the events along it. The simplest way to simulate such an uncorrelated waiting times trajectory is to draw a random time out of $\phi_{on}(t)$, which determines the time duration the process stays in the *on* state, and then to draw a random time out of $\phi_{off}(t)$, which determines the time duration the process stays in the *off* state. By repeating this procedure, a time-series is constructed. Such a two-state process in which the *on* and *off* waiting times are drawn randomly and independently out of the corresponding PDFs is called here a two-state semi-Markov (TSSM) process (Fig. 2F). By construction, all the information in a TSSM process is contained in the waiting time PDFs. This means that any function, see functions (*a*)-(*d*) above, calculated from its two-state trajectory is given in terms of the waiting time PDFs. When a very complex kinetic scheme generates an uncorrelated waiting times trajectory, we say that it is reducible to a TSSM scheme, because of the equivalence of its two-state trajectory to that of a TSSM process. This implies that information about the connectivity between substates within states of reducible schemes is lost. Moreover, it follows that reducible schemes with the same waiting time PDFs cannot be discriminated by the trajectory analysis.

An indication for the lack of correlations in the trajectory is the factorization of $\phi_{x,y}(t_1,t_2)$ $x,y = on, off$, into the product of $\phi_x(t_1)$ and $\phi_y(t_2)$ for *every* $x, y = on, off$,

$$\phi_{x,y}(t_1,t_2) = \phi_x(t_1)\phi_y(t_2) \quad ; \quad x,y = on, off . \tag{2}$$

In principle, there are two possible scenarios that lead to this: (*i*) when the scheme is symmetric due to a special choice of the system parameters, and (*ii*) when the scheme possesses a special connectivity between its *on* and *off* substates (as discussed below, combination of special connectivity and symmetry can lead to the same result). We focus hereafter on characterizing the topology of reducible schemes regardless of the system parameters [case (*ii*)]. For this, we write $\phi_{x,y}(t_1,t_2)$ as,

$$\phi_{x,y}(t_1,t_2) = \sum_{n,m,l} W_n^x f_{mn}^x(t_1) p_{lm} F_l^y(t_2) \quad ; \quad x,y = on, off . \tag{3}$$

Here, $p_{lm}$ is the probability that an event starts at substate *l* of state *y* when the previous event in state *x* terminated at substate *m* of the other state. For example, when $x \neq y$, $p_{lm} = \delta_{lm}$, where $\delta_{ij}$ is the Kronecker delta. According to Eq. (3) and Eq.

(1) that gives two possible definitions for $\phi_z(t)$, a kinetic scheme is reducible, namely, Eq. (2) holds, when $p_{lm} = W_l^y$ for every $x, y = on, off$, leading to,

$$\phi_{x,y}(t_1,t_2) = \sum_{n,l} W_n^x F_n^x(t_1) W_l^y F_l^y(t_2) = \phi_x(t_1)\phi_y(t_2). \tag{4}$$

Although Eq. (4) is not the only option for which Eq. (2) holds, it is the only option for a scheme to be reducible just due to its topology. To characterize this topology, we define a special substate called a gateway substate. A gateway substate is one for which all events in a state either start from, type 1 (Fig. 2E), or terminate at, type 2 (Fig. 2D). A gateway substate, say substate $N$, of type 1 in state $x$ means that,

$$W_n^x = \delta_{Nn}, \tag{5}$$

and, thus, leads to

$$\phi_{z,z}(t_1,t_2) = \phi_z(t_1)\phi_z(t_2) \qquad ; \qquad z = on, off, \tag{6}$$

and,

$$\phi_{y,x}(t_1,t_2) = \phi_y(t_1)\phi_x(t_2) \qquad ; \qquad y \neq x. \tag{7}$$

The factorization of $\phi_{x,x}(t_1,t_2)$ immediately follows from Eq. (3) when using Eq. (5), where the factorization of $\phi_{y,x}(t_1,t_2)$ for $y \neq x$ follows from $p_{lm} = \delta_{lm}\delta_{Nl}$ for this case, because that all events in state $x$ must start at substate $N$. The factorization of $\phi_{y,y}(t_1,t_2)$ for $y \neq x$, meaning $p_{lm} = W_l^y$, occurs because the next event in state $y$ occurs after an event in state $x$ that starts always from substate $N$ is terminated, which means that the hitting probabilities of the irreversible $x$ process are the $p_{lm}$'s, and these are the same for every $y$ state cycle. A similar situation occurs for a gateway substate of type 2 in state $x$, that is, Eq. (5) holds, and in Eq. (6) we substitute $x$ and $y$. However, one gateway substate of type 1 in state $x$ is not sufficient for the factorization of $\phi_{x,y}(t_1,t_2)$ for $x \neq y$, which can be written for this case as,

$$\phi_{x,y}(t_1,t_2) = \sum_m f^x_{mN}(t_1) F^y_m(t_2) \quad ; \quad y \neq x. \tag{8}$$

There are two possibilities that lead to the factorization of Eq. (8) due to the scheme topology: when $f^x_{mN}(t_1) = f^x_{MN}(t_1) W^y_m$, namely, when there is another gateway substate (M) in state x, now of type 2, or when $F^y_m(t_2) = F^y_M(t_2)\delta_{Mm}$, namely, when there is a gateway substate (M) of type 1 in state y. If we start from a type 2 gateway substate in state x, then we should check when $\phi_{y,x}(t_1,t_2)$, which is given by,

$$\phi_{y,x}(t_1,t_2) = \sum_{n,m} W^y_n f^y_{mn}(t_1) F^x_m(t_2) \quad ; \quad y \neq x, \tag{9}$$

factorizes. This happens when $f^y_{mn}(t_1) = f^y_{Mn}(t_1) W^x_m$, namely, when there is a type 2 gateway substate in state y (M), or when $F^x_m(t_2) = F^x_M(t_2)\delta_{Mm}$, namely, when there is another gateway substate (M) in state x, now of type 1. Note that if state y is symmetric in the sense that $F^y_m(t_2) = F^y(t_2)$, then Eq. (8) factorizes, meaning that a combination of special topology and symmetry can lead to reducible schemes as well.

Thus, we have shown that a scheme is reducible due to its topology when it possesses at least one of the following possibilities: (A) two gateway substates of different types in the same state (Fig. 2C), and (B) & (C) two gateway substates of the same type, either type 1 (Fig. 2E) or type 2 (Fig. 2D), in different states. Note that an equilibrium-reached scheme is reducible due to its topology if and only if it possesses a gateway substate of both types. Summarizing the above findings we state that the classes of schemes that fulfill Eq. (1) regardless of the system parameters are those schemes for which each *on* (*off*) event along the trajectory has the same initial probabilities among the *on* (*off*) substates as the previous *on* (*off*) events

**Concluding Remarks**

Single molecule two-state trajectories supply the possibility of obtaining detailed information about the underlying mechanism of the studied process. In a previous work [O. Flomenbom, J. Klafter, and A. Szabo, submitted (2004)] we showed that when the waiting times along the trajectory are uncorrelated, all the information in it is contained in the waiting time PDFs. We called the schemes that produced uncorrelated waiting times trajectory reducible. In this work, we characterized and proved the topologies that lead to reducible schemes. A scheme is reducible when it is symmetric, or possesses a specific connectivity between its *on* and *off* substates (combination of special connectivity and symmetry can lead to the same result). The topology that leads to reducible schemes includes: (*A*) two gateway substates of different types in the same state (Fig. 2C), and (*B*) & (*C*) two gateway substates of the same type, either type 1 (Fig. 2E) or type 2 (Fig. 2D), in different states. Due to this topology, that is, due to this specific connectivity between the scheme substates of different states, it is not possible to obtain information about the connectivity of the scheme substates of the same state. It thus follows that reducible schemes with the same waiting time PDFs cannot be discriminated by the analysis of a trajectory.

Other details about the topology of schemes, as well as the analysis of trajectories from *irreducible* schemes, are given elsewhere[36].

**Acknowledgments**

We thank A. Szabo, A. M. Berezhkovskii and I. Gopich for fruitful discussions.

**Figure Captions:**

**Figure 1** A trajectory of an observable that fluctuates between two values (*on* and *off*) as a function of time.

**Figure 2 A-E** A set of reducible kinetic schemes, and a TSSM scheme (**F**), characterize only by the waiting time PDFs $\phi_{on}(t)$ and $\phi_{off}(t)$. **A** An *n* uncoupled *off* substates connected to one *on* substate. The dashed line represents the *off* substates that are not shown. **B** An *n* coupled *off* substates with one o*n* substate scheme. **C** A reducible scheme with two gateway substates that are in the same state (the *on* state). The bolded pentagons with full lines (connectors) stand for a region with any complex network of connections within a state. The dashed arrow stands for a set of connections between many *off* substates and one *on* substate, and the dashed–dotted arrow stands for a set of connections between one *on* substate and many *off* substates. **D-F** When the gateway substates in both the *on* and the *off* states are of type 1 (**D**), or of type 2 (**E**), the scheme is reducible to a TSSM scheme (**F**).

FIG 1

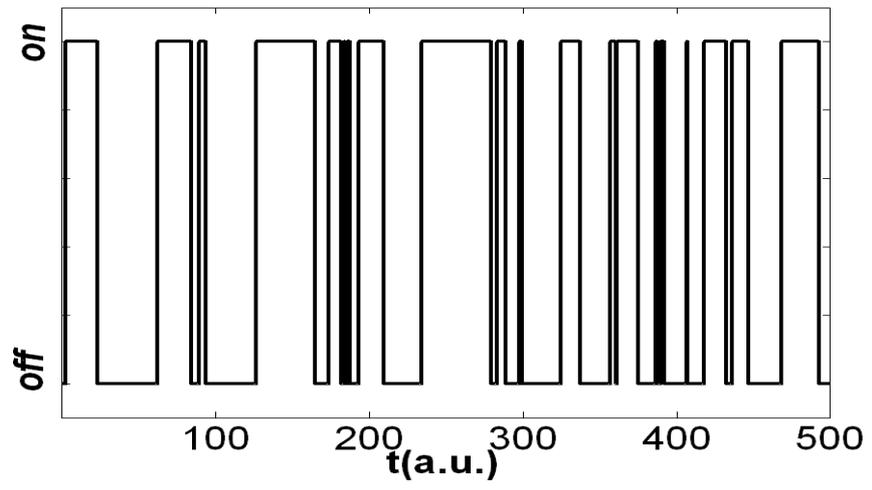

FIG 2

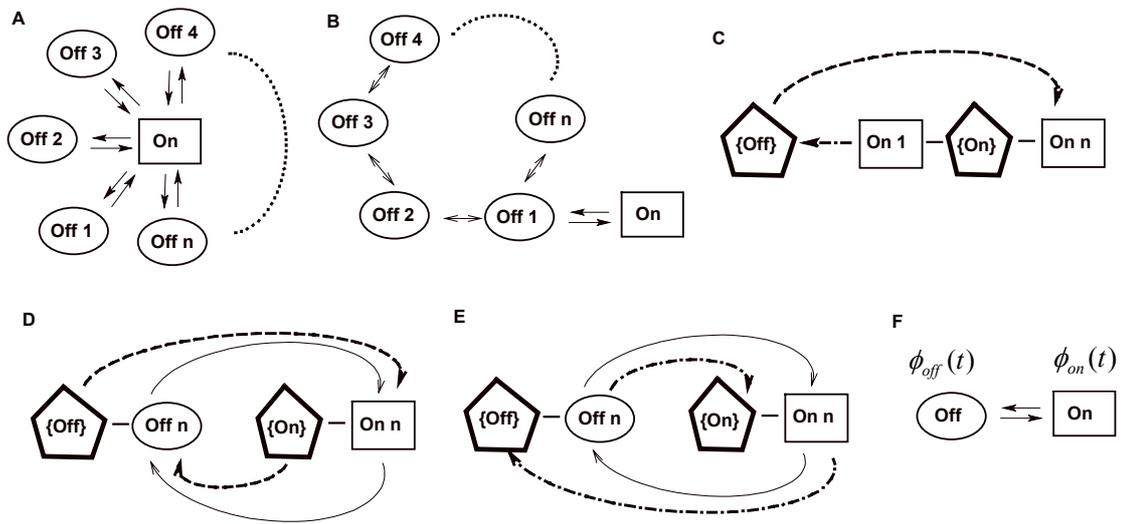